# Correlation of multiple sclerosis incidence in the UK 1990-2010 with geomagnetic disturbances

Seyed Aidin Sajedi [1 *], Fahimeh Abdollahi [2]

1. Department of Neurology, Golestan Hospital, Ahvaz Jundishapur University of Medical Sciences, Ahvaz, Iran

2. Department of Internal Medicine, Golestan Hospital, Ahvaz Jundishapur University of Medical Sciences, Ahvaz, Iran

* Correspondence: Dr. Seyed Aidin Sajedi, Department of Neurology, Golestan Hospital, Farvardin Street, Ahvaz, Iran   email: dr.sajedy@gmail.com

A commentary on

Incidence and prevalence of multiple sclerosis in the UK 1990–2010: a descriptive study in the General Practice Research Database.

*By Mackenzie IS, Morant SV, Bloomfield GA, MacDonald TM and O'Riordan. J Neurol Neurosurg Psychiatry (2013) doi: 10.1136/jnnp-2013-305450.*

We read with the great interest the recent article by Mackenzie et al. reporting incidence and prevalence of multiple sclerosis (MS) in the UK 1990-2010 (1). Beyond the various advantages of such descriptive studies about the epidemiology of MS, the most important and useful aspect of this study, in our opinion, is determination of long-time MS incidence trend. However, MS epidemiology is studied frequently to find clues about possible environmental risk factors, long-time reports of its annual incidence for consecutive years are relatively rare. While, such incidence trend reports provide valuable opportunity to test the ability of risk factor hypothesis in describing the alterations of disease incidence, it seems that one of the causes of the rarity of such reports in publication may be the fact that often-cited hypotheses like vitamin D deficiency hypothesis and chronic cerebrospinal venous insufficiency (CCSVI) cannot explain MS incidence trends.

Nevertheless, there is a newcomer that may provide some explanation. Last year, we framed and tested a new hypothesis for MS based on the biological effects of geomagnetic disturbances

(GMD) (2). By a vast ecological study, we found that GMD hypothesis has the greatest ability to explain MS prevalence distribution and also can provide answers for issues such as its relation to month of birth, immigration, the cause of gradual attenuation of MS prevalence gradient, and the historical trend of its global incidence.

Recently, we tried to test the ability of GMD hypothesis in describing MS incidence alteration in specific locations and time periods. A preliminary test on consecutive report of previously published data of annual MS incidence in two locations during the 23rd solar cycle (1996-2008) showed that there may be significant correlation among MS incidence and GMD (3). Nevertheless, we were very keen to find data with a better acquisition method and registry, from areas near or under geomagnetic 60 degree latitude (GM60L), which experience the greatest amount of GMD, to evaluate the probable existence of such correlation. Incidence data of the UK 1990-2010 provided such opportunity. Accordingly, we calculated annual averages of planetary A index ($A_P$), a main GMD index retrieved from Goddard spaceflight center-space physics data facility for the period of 1985-2010 (figure 1A), then, possible lead-lag relationships among $A_P$ and MS incidence were evaluated by means of cross-correlation analysis (4) for lags between -3 to +3 years. Approximate data of annual MS incidence of the UK were retrieved from figure 3A of Mackenzie's et al. report (1).

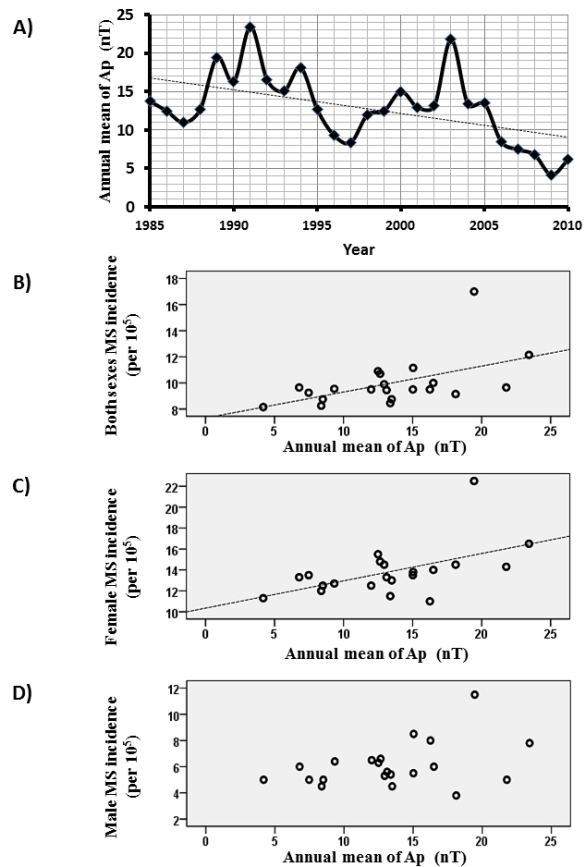

Figure 1. A) Annual average of $A_P$ index. B to D) Scatter plots of correlation among annual average of $A_P$ with annual MS incidence of the UK 1990-2010 (1).

Results of cross-correlation analyses showed that with a one year delay, there may be a significant correlation between annual average of $A_P$ with both sexes MS incidence, female MS incidence, and male MS incidence. After correction of one year delay for the time series of Ap , a bivariate Pearson's correlation analysis showed that there is a moderate significant correlation between average of $A_P$ (1989-2009) and both sexes MS incidence of 1990-2010 (r= 0.52, P=0.016)(figure 1B). However, sex specific analyses showed that like our previous study (3), this significant correlation is resulted from correlation of $A_P$ with MS incidence of females (r=0.53, P=0.013) but not with males MS incidence (r=0.39, P=0.078)(figure 1C and 1D).

The observed one year delay among GMD and MS incidence was the same as observed in our previous study on western Greece data (3). We believe that this delay, however, can be considered as a clue in favor of probable causal relationship among risk factor and the disease, but cannot be regarded as a real delay-time for observing the effect of this probable risk factor. In our opinion, this delay is mainly resulted from the inevitable delay due to the need to meet the criteria of "dissemination in time" in the diagnostic criteria of MS, and the fact that all MS incidence reports are based on the time of diagnosis confirmation and not the time of the first attack of disease.

As Mackenzie et al indicated, they could not explain the cause of decreasing MS incidence trend in the UK, while, by using GMD hypothesis this trend can be explained. In the time of framing GMD hypothesis, we discussed how it can explain historical MS incidence trend (2). Except than a period of 1960-1970, GMDs had an increasing overall trend since 1900 up to about 1990 due to increasing solar activities. But, after 1990 and especially since 2004 the overall trend has been decreasing. If we regard GMD as an important MS risk factor, then the observed decreasing trend in MS incidence would not be surprising. In addition, the cause of significant higher incidence and prevalence of MS in Scotland can be explained easily, as it is located exactly under the hot line of GM60L.

This evaluation again indicated that we need a substantial revision in the method of reporting MS incidence in future epidemiological researches. Important message of this finding for MS investigators, especially in high latitudes near GM60L that experience the most GMDs, is to provide long-term reports of MS incidence with higher resolution, i.e. at least monthly data of MS incidence based on the exact time of disease onset, not just the time of diagnosis confirmation.

Providing data of the exact time of relapses also seem to be useful and maybe more practical due to the awareness and sensitivity of patients to report promptly any new signs or symptoms to their physicians. Then, it would be possible to conduct superposed epoch analyses to investigate actual abilities and validity of any potential risk factor and related hypothesis. Clarifying this fact especially for GMD hypothesis may be quiet valuable, not only for better understanding of MS pathophysiology, but also for prevention of relapses by modifying treatments just before increasing exposure to the risk factor; due to the fact that observational stations in the Earth orbit let us to predict significant GMD before occurrence.